\documentclass[journal]{IEEEtran}
\usepackage{amsmath}
\usepackage{algorithm}
\usepackage{algpseudocode}
\usepackage{array}
\usepackage{url}
\usepackage{graphicx}
\usepackage{cite}
\usepackage{amssymb}
\usepackage[dvipsnames]{xcolor}
\usepackage{multirow}
\usepackage[table]{xcolor}
\usepackage{colortbl}
\usepackage{booktabs}
\usepackage{hyperref}
\usepackage{multirow}
\usepackage{xcolor}
\usepackage{bm} % For bold math symbols
\usepackage{makecell}
\usepackage{array}
\usepackage{makecell}
\usepackage{graphicx}
\usepackage[caption=false,font=footnotesize]{subfig} % IEEEtran-friendly
\graphicspath{{Figure/}}

\usepackage{booktabs}
\usepackage{tabularx}
\usepackage{array}
\newcolumntype{L}{>{\raggedright\arraybackslash}X}  % left-aligned, wraps
\newcolumntype{C}{>{\centering\arraybackslash}X}    % centered, wraps
\newcolumntype{R}{>{\raggedleft\arraybackslash}X}   % right-aligned, wraps

\newcommand{\real}{\mathbb{R}}

\begin{document}

\title{A Data-Driven Method to Identify Major Contributors to Low-Frequency Oscillations}

\author{
    Youhong~Chen,~\IEEEmembership{Member,~IEEE,} 
    Debraj~Bhattacharjee,~\IEEEmembership{Student Member,~IEEE,} 
    and~Balarko~Chaudhuri,~\IEEEmembership{Fellow,~IEEE}%
    \thanks{The authors are with the Department of Electrical and Electronic Engineering, Imperial College London, London, SW7 2AZ, UK (e-mail: youhong.chen24@imperial.ac.uk).This work is supported by the Engineering and Physical Science Research Council, UK [grant number EP/Y025946/1]}
}

\maketitle

\begin{abstract}
We present a purely data-driven method to pinpoint generation plants that significantly contribute to poorly damped oscillations as part of post-event analysis. First, Extended Dynamic Mode Decomposition (EDMD) is applied on PMU data from the point of interconnection (POI) of the plants to obtain the finite-dimensional Koopman operator. Then, modal analysis is performed on a reduced-order Koopman operator to extract spatio-temporal patterns. The data-driven eigenvalues and eigenvectors quantify each plant’s contribution to critical oscillatory modes without requiring any system model information. We demonstrate the effectiveness of this method through simulated case studies on modified IEEE 39-bus and WECC 179-bus test systems by benchmarking the data-driven results against ground-truth models. Its performance is further validated using PMU data from real oscillation events in the ISO-New England system. This data-driven method offers a practical tool for both planning-stage simulations and post-event analysis of real oscillation events, enabling effective mitigation.
\end{abstract}

\section{Introduction}
\IEEEPARstart{P}{oorly} damped low-frequency oscillations (e.g., inter-area oscillations) have always been a concern for system operators. As power systems transition towards a high share of inverter-based resources (IBRs) such as wind, solar photovoltaics, grid-scale battery energy storage etc., new stability challenges are emerging rapidly. This includes poorly damped sub-synchronous oscillations (SSOs) caused by adverse interaction among IBRs and the network. Recent IBR-induced SSO events \cite{TPWRS23_Fan_SSO_realworld_events,PEMag24_Modi_Real_world_SSO_events} outline the scale of the challenge. Unforeseeable risks of SSO often force the system operators to adopt a conservative approach by limiting the maximum share of IBRs connected to the grid at a given point. 

% Oscillations in power systems are broadly classified as forced or natural. Forced oscillations occur when a component injects energy at an oscillatory frequency and persist as long as the external forcing continues \cite{esig_oscillations_2024}, with the device receiving the forcing signal,or the component dissipating energy in energy-based methods, typically regarded as the source \cite{Maslennikov2021}. Natural oscillations arise from interactions among many elements that collectively produce poor damping \cite{esig_oscillations_2024}. In this category, no single component can be identified as the source or "bad actor" because the oscillation results from complex energy exchanges among multiple devices. Consequently, mitigation focuses on identifying the components whose damping adjustments can most effectively suppress the oscillation \cite{Estevez2025}. Accordingly, this paper focuses on identifying the significant contributors to poorly damped oscillations rather than locating a single source.

Oscillations are broadly classified as forced or natural. Forced oscillations arise when a generation plant injects energy at a specific frequency that persists until the forcing continues \cite{esig_oscillations_2024}.  The generation plant subject to the forcing signal is typically considered the source or the `bad actor' \cite{Maslennikov2021}. Natural oscillations, by contrast, result from adverse interactions among more than one elements \cite{esig_oscillations_2024}, with no single `bad actor.' Mitigation, therefore, targets generation plant(s) with major contribution to the oscillation \cite{Estevez2025}. Hence, this paper focuses on identifying significant contributors to poorly damped oscillations.

Identifying root-cause, i.e, the significant contributors to poorly damped oscillations, is critical for effective mitigation. However, this task is becoming increasingly challenging with high IBR penetration. This is mainly because the dynamic behavior of an IBR-dominated power system is largely dictated by proprietary control algorithms embedded in vendor-specific black-box IBR models. This lack of transparency makes traditional model-based stability analysis impractical as IBR internal control structures and parameters are unavailable to the system operators. As a result, it is essential to develop a data-driven approach that can quantify the relative contribution of each power plant, either SMs or IBRs, and also extend to other key assets such as HVDC links, FACTS devices, and large loads, in order to enable effective mitigation of poorly damped oscillations. 

% This has to be based on time-synchronized measurements (real or simulated) from the points of connection (PoC) of the IBRs.

Techniques based on Dissipating Energy Flow (DEF)  have proven to be effective in identifying the source(s) of forced oscillations, particularly those originating from synchronous resources \cite{Maslennikov2021}. However, extending DEF to IBR-dominated systems has been particularly challenging \cite{Fan_Slava_23_DEF_IBR-drivenSSO} until a modified SSO-CDEF method \cite{Estevez2025} was proposed recently. DEF-based methods generally rely on extensive PMU coverage to trace dissipating energy flows, and limited coverage can hinder accurate tracing of the source. The magnitude (or amplitude) of SSO at the point of interconnection (POI) of a generation plant may be an indicator of its degree of contribution in SSO \cite{dong2023analysisnovember212021} but is not always reliable, as we demonstrate in this paper. This necessitates other reliable data-driven algorithms.

Dynamic Mode Decomposition (DMD) is a data-driven, model-free technique for capturing the spatiotemporal behavior of complex systems. This is widely used in the fluid mechanics community for modal analysis, actuator selection, and sensor placement \cite{schmid2022dynamic, bhattacharjee2020data, sashittal2021data}. Owing to its flexibility and low computational cost, DMD has been applied in power systems to identify system dynamics with Kalman filtering for state estimation and model predictive control \cite{8384030,9780619}, and to assess nonlinear system's observability within a Koopman framework \cite{10219065}. For oscillation studies, DMD enables data-driven modal analysis to detect critical modes efficiently and accurately from large PMU datasets \cite{6982236,9151361,10328697,9383277}. Prior work shows that DMD analysis can efficiently process and analyse large datasets from multiple measurement sources compared with the Prony method\cite{6982236}. The work in \cite{9151361} employs a DMD-based approach to effectively identify and reconstruct poorly damped mode oscillations from real-world inter-area oscillation events using shift-stacking technique. However, \cite{9151361} stops at mode identification without quantifying the participation of individual components. To address this, \cite{10328697} developed a DMDc-based source location technique by substituting the control input with initial measurement states. They propose a contribution factor index derived from DMD-based mode shapes to quantify the contribution of each bus in critical modes and, thereby, determine oscillation sources. However, this approach relies on bus voltage magnitudes and phase angles only, which might be insufficient to capture the nonlinear power system dynamic behavior adequately. Moreover, the methodology was demonstrated only for single-mode oscillation events, whereas in practical power systems, poorly damped multi-modal oscillations are common.

We demonstrate that Extended Dynamic Mode Decomposition (EDMD), a variant of DMD that captures the nonlinear dynamics of the underlying system, can accurately identify the major contributors in terms of their relative participation in poorly damped low-frequency oscillation. While an appropriate choice of observables remains a relatively open problem in the EDMD space, we show that active and reactive power,
which are polynomial and trigonometric functions of voltage
and current phasors obtained from PMUs, produce promising
results. Notably, this approach avoids the need for 1) data stacking that increases the dimensionality of the problem \cite{9151361} or 2) additional post-processing \cite{10328697}. 

The main contributions of this paper are:

\begin{enumerate}
    % \item  A data-driven method is developed to identify which power components have dominant participation in poorly damped low-frequency oscailltion. The accuracy of the method is validated by comparing the rank order of the data-driven participation factor against those obtained from the actual linearized state-space model.

    % \item  An extend dynamic mode decomposition (eDMD) method with an appropriate choice of observables is proposed to improve the accuracy of data-driven participation factor analysis of SSO. This advances the state-of-the-art at applying DMD for data-driven root cause analysis of IBR-induced SSO.

    \item An EDMD-based data-driven method is developed to identify generation plants with dominant contribution in poorly damped low-frequency oscillations. The method, grounded in Koopman operator theory, is shown to be effective for root-cause analysis of both natural and forced oscillation events without requiring extensive PMU coverage.

  \item A systematic method that integrates FFT, EDMD, and data-driven Koopman modal analysis is established to accurately determine dominant oscillation modes and their major contributors. A reduced-order Koopman operator is used to minimize computational burden.

  % \item An appropriate choice of observables for EDMD enhances the ability to capture nonlinear system dynamics and thereby improves the accuracy of participation factor analysis in low-frequency oscillations.

\end{enumerate}

% \subsection{Background and Motivation}

% \subsection{Related Work}

% \subsection{Original Contributions}

\section{Extended Dynamic Mode Decomposition (EDMD)}

% Dynamic Mode Decomposition (DMD) is a data-driven approach for capturing the spatiotemporal behavior of complex systems such as power grids. The DMD method finds its origins in the fluid mechanics community, where it has been used for a wide variety of purposes, such as modal analysis, actuator selection, sensor placement~\cite{schmid2010dynamic, schmid2022dynamic, bhattacharjee2020data, sashittal2021data}. Owing to its flexibility and relatively low computational requirements, the application of DMD has been explored in power systems \cite{8384030, 9780619, 10219065}. It is closely related to the Koopman formalism for dynamical systems, which allows the decomposition of a set of observables obtained from a complex system into linearly superimposed simple processes. In particular, the evolution of the system's state and related observables is governed by the Koopman operator~\cite{koopman1932dynamical}, which is linear but infinite-dimensional. The \textit{slow} subspace of this operator can be used to capture long-term dynamics of observables and can serve as a surrogate of the otherwise infinite-dimensional operator. Once the Koopman operator is obtained from data, the state of the underlying system can be reconstructed using its eigenvalues, eigenfunctions, and models. 
The EDMD is closely related to the Koopman formalism for dynamical systems. This allows the decomposition of a set of observables obtained from a complex system into linearly superimposed simple processes. In particular, the evolution of the system's state and related observables is governed by the Koopman operator~\cite{koopman1932dynamical}, which is linear but infinite-dimensional. The \textit{slow} subspace of this operator can be used to capture long-term dynamics of observables and can serve as a surrogate of the otherwise infinite-dimensional operator. Once the Koopman operator is obtained from measurement data, the dynamics of the underlying system can be analyzed using Koopman eigenvalues, eigenfunctions, and modes. 

Since the Koopman operator is the central piece of this formulation, we briefly recall its definition and related properties~\cite{williams2015data}. To this end, consider a discrete-time dynamical system of the form
\begin{align}
    x_{k + 1} = F(x_k),
    \label{eqn: nls}
\end{align}
where $x \in \mathcal{M} \subseteq \real^n$, $k \in \mathbb{Z}$, and $F: \mathcal{M} \rightarrow \mathcal{M}$. Then the Koopman operator $\mathcal{K}$ acts on a space of functions of observables $\psi \in \mathcal{F}$ such that
\begin{align}
    \mathcal{K} \psi = \psi \circ F,
    \label{eqn: Koopman_composition}
\end{align}
where $\psi: \mathcal{M} \rightarrow \mathbb{C}$ and $\circ$ denotes the composition of $\psi$ with $F$. In effect, this transformation is equivalent to a new infinite-dimensional linear system that evolves in the form
\begin{align}
    \psi_{k+1} =  \mathcal{K} \psi_k.
    \label{eqn: Koopman_linear}
\end{align}

\noindent The dynamical systems~\eqref{eqn: Koopman_composition} and~\eqref{eqn: Koopman_linear}, defined by the operators $F$ and $\mathcal{K}$, respectively,  are two distinct parameterizations of the same fundamental behavior.  As shown in~\cite{mezic2005spectral}, the system state at all times in the future can be obtained by evolving $x$ through $F$ or by using the relation
\begin{align}
    F(x) = \sum_{j = 1}^{N_j} v_j (\mathcal{K}\phi_j) (x) =\sum_{j = 1}^{N_j} \mu_j v_j \varphi_j (x),
    \label{eq: KMD}
 \end{align}
where $\mu$ represents a specific eigenvalue associated with the Koopman operator, whereas $\varphi$ and $v$ represent the corresponding eigenfunction (see~\cite{uhlenbeck1976generic} for more details) and mode, respectively. Since working with an infinite-dimensional but linear operator is simpler for design and analysis as compared to a finite-dimensional but nonlinear one, approximation of the Koopman operator from measured data has attracted a lot of interest in recent times.  For nonlinear systems, in particular,  the approximation of the Koopman operator and the subsequent reconstruction of the original state hinge on the choice of observables that the designer selects. The procedure to compute this approximation directly from measured data is known as Extended Dynamic Mode Decomposition (EDMD)~\cite{williams2015data}.

 The dataset to be used should consist of successive ``snapshot" pairs or observables. This set of data is then used with a dictionary of functions that spans the subspace of such observables. Consider a dataset of snapshot pairs $\{ (x_m, y_m) \}_{i = 1}^M$ such that
\begin{equation}
%\begin{split}
      X :=  \begin{bmatrix}
        | & | &        & | \\
        x_{1} & x_{2} & \hdots & x_{M} \\
        | & | &        & | 
      \end{bmatrix},
      Y :=  \begin{bmatrix}
        | & | &        & | \\
        y_{1} & y_{2} & \hdots & y_{M} \\
        | & | &        & | 
      \end{bmatrix},
%\end{split} 
\label{eqn:dataXY}
\end{equation}
where $x_i, y_i \in \mathcal{M}$ and $y_i = F(x_i)$ for all $i = 1, \hdots, M$. In addition, consider a dictionary $\mathcal{D} = \{ \psi_1, \hdots, \psi_{n_d} \}$, such that $\psi_i \in F$ for $i = 1,\hdots, N_d$. Finally, we also define the vector-valued function $\Psi: \mathcal{M} \rightarrow \mathbb{C}^{1 \times n_d}$
\begin{align}
    \Psi(x) = \big[ \psi_1(x) \, \,  \psi_2(x) \, \, \hdots \, \, \psi_{n_d}(x) \big].
\end{align}

\noindent A finite-dimensional representation of the Koopman operator can be computed as
\begin{equation}
    K \triangleq G^{\dagger} H, 
    \label{eq:koopmandef}
\end{equation}
where the matrices $G$ and $H$ are given by
\begin{align}
    G = \frac{1}{M} \sum_{j = 1}^M \Psi(x_j)^* \Psi(x_j),
    % \label{def_G}
    \quad
    H = \frac{1}{M} \sum_{j = 1}^M \Psi(x_j)^* \Psi(y_j),
    \label{def_H}
\end{align}
and $G^{\dagger}$ represents the pseudo-inverse of the matrix $G$. 

Until now, we presented the standard EDMD procedure to approximate the Koopman operator in a finite-dimensional subspace. 
%\hlText{ An equivalent formulation of the Koopman operator $K$ develop in \cite{Klus2016}, which is more convenient for computing the data-driven participation factors, is adopted as in \cite{Netto2019} and expressed as
% \begin{equation}
%     K^\top \triangleq H G^{\dagger}, 
%     \label{eq:koopmandef_v2}
% \end{equation}
%where H and G are the snapshot matrices defined in \eqref{def_G} and \eqref{def_H}.} 
% It was shown in~\cite{Klus2016} that the operator $M_K = K^\top$ can be used for computing participation factors directly from data.
An equivalent formulation of the Koopman operator $K$, proposed in \cite{Klus2016}, where
\begin{align}
    M_K=K^\top,
    \label{MK_def}
\end{align}
is adopted in this work. This formulation was also used in \cite{Netto2019} because it is more convenient for computing data-driven participation factors, and is followed here for consistency.

Therefore, if $\xi_i$ is the $i-$th left eigenvctor of $M_K$, with corresponding eigenvalue $\mu_i$, the approximation of this eigenfunction through EDMD is given by 
\begin{align}
    \varphi_i (x) = \xi_i \Psi^\top (x) .
    \label{eigfundef}
\end{align}

Finally, to obtain the Koopman modes corresponding to the full-state observable $x_k$, we define $B \in \mathbb{R}^{n \times n_d}$ such that
\begin{equation}
    x = B \, \Psi^\top (x).
    \label{eq:B_def}
\end{equation}
As an intermediate step, we define a vector-valued function $\varPhi: \mathcal{M} \rightarrow \mathbb{C}^{n_d \times 1}$
\begin{align}
     \varPhi(x) &= [\varphi_1(x) \; \varphi_2(x) \; \dots \; \varphi_{N_d}(x)]^\top,
     \label{eq:phi_def}
\end{align}
which can also be expressed as
\begin{align}
    \varPhi(x) &= \Xi \Psi^\top (x)\, ,
     \label{eq:eigfun_def2}
\end{align}
where  $\Xi=[\xi_1^\top \; \xi_2^\top \; \dots \; \xi^\top_{n_d}]^\top \in \mathbb{R}^{n_d \times n_d}$ is the matrix of left eigenvectors associate with $M_K$. We define a matrix $\Phi\in \mathbb{R}^{n \times n_d}$ as
\begin{equation}
\Phi = B \Xi^{-1},
\label{eq:defKoopman_mode}
\end{equation}
where the column vector $\phi_i$ of $\Phi$ corresponds to the Koopman modes. Therefore, combining~\eqref{eq:B_def} and ~\eqref{eq:eigfun_def2}, we obtain
\begin{align}
    x &= \sum_i \varphi_i\phi_i 
    \label{eq:koopman_modes}
\end{align}

% Finally, to obtain the Koopman modes corresponding to the full-state observable $x_k$, we define a matrix $B \in \mathbb{R}^{N_d \times N}$ as
% \begin{equation}
%     x = B^\top \, \Psi(x)^\top = (\Psi(x) B)^\top,
%     \label{eq:B_def}
% \end{equation}
% where $N$ is the number of scalar-valued observables. As an intermediate step, we define a vector-valued function $\Phi: \mathcal{M} \rightarrow \mathbb{C}^{1 \times N_d}$
% \begin{align}
%      \Phi(x) &= [\phi_1(x) \; \phi_2(x) \; \dots \; \phi_{N_d}(x)],
%      \label{eq:phi_def}
% \end{align}
% which can also be expressed as
% \begin{align}
%     \Phi(x) &= \Psi(x)\, \Xi,
%      \label{eq:eigfun_def2}
% \end{align}
% where $\Xi=[\xi_1 \; \xi_2 \; \dots \; \xi_{n_k}]$ is the matrix of eigenvectors whose inverse is denoted by $K$. 
% The matrix  $K$ can further be written out as
% \begin{equation}
%     \Xi^{-1} = L^* = [\ell_1 \; \ell_2 \; \dots \; \ell_{N_d}]^*,
%     \label{eq:Xi_inverse}
% \end{equation}
% where $\ell_i$ denotes the $i$-th left eigenvector of $K$ associated with eigenvalue $\mu_i$. These eigenvectors are then scaled so that $\ell_i^* \xi_i = 1$.

% \noindent Therefore, by combining~\eqref{eq:B_def},~\eqref{eq:eigfun_def2} and~\eqref{eq:Xi_inverse}, we obtain
% \begin{align}
%     x &= \hat{L} \, \Phi(x)^\top = \sum_{k=1}^{N_k} \hat{\ell}_k \, \phi_k(x), \nonumber \\
%     \hat{L} &= [\hat{\ell}_1 \; \hat{\ell}_2 \; \dots \; \hat{\ell}_{N_d}] = (L^* B)^\top
%     \label{eq:koopman_modes}
% \end{align}
% where $\hat{\ell}_i = (\ell_i^* B)^\top$ is the $i$-th Koopman mode.

In summary, the Koopman eigenfunctions $\varPhi = \Xi \Psi^\top$ are approximated by the left eigenvectors of $M_K$, while the Koopman modes $\Phi$ are obtained as $\Phi = B\Xi^{-1}$, where $\Xi^{-1}$ contains the right eigenvectors of $M_K$ and $B$ is the coefficient matrix defined in \eqref{eq:B_def}. Unlike applications where EDMD is primarily used for signal reconstruction, this work applies EDMD to identify critical oscillation modes and their major contributors. Accordingly, the matrix $B=I$, so the Koopman modes reduce to the right eigenvectors of $M_K$, i.e, $\Phi =\Xi^{-1}$.

% It should also be noted that the accuracy of Koopman modal analysis using EDMD depends on the choice of the observable library. Selecting a set of observables that spans a true Koopman-invariant subspace is generally difficult and remains an open research question. To address this challenge, the Koopman operator is projected onto a reduced-dimensional subspace to preserve the key dynamic information while reducing computational burden. The details of this projection are discussed in Section III.

\section{Proposed Methodology}
\label{sec:Methodology}

\subsection{Selection of Observables}
\label{subsec:Sel_of_Obser}

The first step of the EDMD approach is to define the a dictionary of functions that lifts the given PMU data into a higher-dimensional subspace $\Psi$, ensuring that the system dynamics are adequately captured. As power systems are inherently nonlinear, it is essential that these nonlinear effects are captured effectively while determining major contributors to poorly dampled oscillations.  Herein we find guidance in~\cite{9917430}, which suggests using polynomial and trigonometric basis functions from the voltage and current phasor data $V \angle \theta_V$, $I \angle \theta_I$ to capture the nonlinear behavior of
a power system. Hence, the active power $P = VI\cos(\theta_V - \theta_I)$ and reactive power $Q = VI\sin(\theta_V - \theta_I)$ represent a typical form of quadratic polynomial expansion of the phasor data $V$, $I$, $\theta_V$, $\theta_I$ from PMUs that can effectively capture the nonlinear dynamics. Following this, $P$ and $Q$ data at the POI of generation plants derived out of measured voltage and current phasors from PMU are used as lifted observables for EDMD in this paper. In principle, the accuracy can be improved by augmenting the observable space with additional nonlinear functions of PMU data. However, it increases the dimension of the Koopman operator $K$ and may introduce spurious modes. A systematic procedure for selecting appropriate observables could enhance accuracy and will, therefore, be a focus of our future work.

\subsection{Filtered Mode of Interest}

One of the key steps in the proposed data-driven framework is to identify the dominant poorly damped oscillatory modes $f_s$, i.e., the modes of interest. Focusing on one mode at a time improves accuracy and is commonly used (e.g., in DEF-based methods). The peaks in the FFT of detrended time-series data with amplitudes exceeding a predefined threshold are classified as significant, and their corresponding frequencies are designated as dominant modes $f_s$. The identified frequency then informs both the selection of the data window and the EDMD truncation order in the later steps. For post-event analysis, the data window is chosen to include at least 5 cycles of the lowest dominant frequency.

% As noted in \cite{Maslennikov2017}, FFT results can be significantly affected by trends in measurement signals, particularly in the low-frequency range. To mitigate this, time-series power flows are used as the primary input to FFT, as they exhibit fewer trends and more accurately reflect the MW oscillations most relevant to system operators.

Once a dominant frequency is identified, the observables are filtered within the frequency range of interest before applying EDMD. Although EDMD can be applied to unfiltered data, bandpass filtering enhances accuracy by isolating the dynamics of interest, especially in cases with multiple oscillatory modes. Filtering also reduces the required truncation order while preserving essential system behavior. In this work, a 4\textsuperscript{th}-order Butterworth bandpass filter is employed with cutoff frequencies $f_c = [0.9, 1.1] f_s$. Zero-phase digital filtering is performed by applying the filter in both forward and reverse directions. To mitigate edge effects, only the central 50\%–80\% of the filtered time-series data is retained and used as input to EDMD.

\subsection{Calculation of Reduced-Order $\Tilde{M}_K$}

The filtered data is processed using EDMD to obtain a finite-dimensional approximation of the Koopman operator $M_K$. While accurate analysis ideally requires observables spanning a Koopman-invariant subspace, in practice only an approximately invariant set $\Psi(x)$ can be identified. Poor choices of observables may introduce spurious eigenvalues~\cite{williams2015data}. These often appear as multiple artificial modes around the true frequency, which reduces accuracy. Additionally, in practical power systems, the dimension of $M_K$ could be often prohibitively large. Therefore, $M_K$ is projected onto a reduced subspace via the truncated singular value decomposition (SVD) using~\eqref{eqn:hatKmatrix}. To compute this reduced-order approximation of $M_K$, we first compute the singular value decomposition of $G$, $G = U \Sigma R^*$. Following this, we truncate these matrices using a truncation order $r$ such that $U_r = U(:, 1:r), \quad \Sigma_r = \Sigma(1:r, 1:r), \quad R_r = R(:,1:r)$, where $U_r \in \real^{n \times r}$, $\Sigma_r \in \real^{r \times r}$, and $V_r \in \real^{r \times r}$. Based on this, the reduced-order Koopman operator $\Tilde{M}_K$ is defined as
\begin{align}
\Tilde{M}_K = U_r^*HR_r \Sigma_r^{-1} ,
\label{eqn:hatKmatrix}
\end{align}
The reduced-order Koopman operator $\Tilde{M}_K$ preserves the slower subspace of the otherwise infinite dimensional Koopman operator. This naturally aligns with our objective of analysing low-frequency oscillations, which typically occur in the range of 0.1--15~Hz. In addition, the reduced operator provides a more accurate representation of the system dynamics at the dominant modes.

A critical step in this process is the selection of the truncation order of $\Tilde{M}_K$. Let $\Sigma = \{\sigma_1, \sigma_2, \ldots, \sigma_{n_s}\}$ denote the set of singular values in descending order. The truncation rank is determined by identifying elbow points in the singular value barplot, using sign changes in the discrete second derivative. 

This method, combined with results from FFT, ensures that all dominant oscillatory modes observed in the spectral domain are retained in the reduced-order model, while extraneous modes are excluded. Next, an eigen decomposition of the matrix $\Tilde{M}_K$, given by
\begin{align}
    \Tilde{M}_K \Tilde{\Phi}  = \Tilde{M} \Tilde{\Phi} ,
    \label{eqn:eigvals}
\end{align}
yields the set of right eigenvectors of the reduced-order matrix $\Tilde{M}_K \in \real^{r \times r}$, and the diagonal eigenvalue matrix $\Tilde{M}\in \real^{r \times r}$. 
The right eigenvectors of the Koopman operator matrix $M_K$ can then be obtained from the reduced eigenvectors $\Tilde{\Phi}\in \real^{r \times r}$ using
\begin{align}
    \hat{\Phi} = U_r\Tilde{\Phi},
    \label{eqn:righteigenvector}
\end{align}
where $\hat{\Phi}\in \real^{n \times r}$ is generally a rectangular matrix.  The corresponding left eigenvectors of $M_K$ are computed using the Moore–Penrose pseudoinverse of $\hat{\Phi}$:
\begin{align}
    \hat{\Xi} = \hat{\Phi}^{\dagger} = \hat{\Phi}^{-1} U_r^{\mathrm{T}} (U_r U_r^{\mathrm{T}})^{-1}.
    \label{eqn:lefteigenvector}
\end{align}
where  $\hat{\Xi}\in \real^{r \times n}$. The $\hat{\Xi}$ and $\hat{\Phi}$ obtained from \eqref{eqn:righteigenvector} and \eqref{eqn:lefteigenvector} will be used in the following section for data-driven participation factor calculation to determine the major contributors to the poorly damped oscillations.

\subsection{Data-Driven Modal Analysis}

The eigenvectors of the reduced-order Koopman operator $\Tilde{M}_K$ 
% is used to perform data-driven modal analysis to identify critical oscillation modes and , $\hat{\Xi}$ and $\hat{\Phi}$, of this reduced-order operator
are projected back to the original higher-dimensional subspace using \eqref{eqn:righteigenvector} and \eqref{eqn:lefteigenvector} to determine major contributors to the oscillation modes. % Let $\Lambda$ be a diagonal eigenvalue matrix, as defined in \eqref{eqn:eigvals}, with diagonal entries $\{\lambda_i\}_{i=1,\dots,n}$.
 Note that the eigenvalues $\mu_i$ are obtained directly from \eqref{eqn:eigvals}, requiring only $\Tilde{M}_K$. These eigenvalues are in the discrete-time domain and can be converted to continuous-time form by
\begin{equation}
    \lambda_i = \frac{\ln(\mu_i)}{\Delta t},
    \label{eq:continuous_eigenvalue}
\end{equation}
where $\lambda_i$ is the continuous-time eigenvalue, $\Delta t$ is the sampling interval, and $\mu_i$ is the discrete-time eigenvalue obtained from EDMD. For each eigenvalue $\lambda_i = \frac{1}{2 \pi}\left({\sigma_i + j\omega_i}\right)$, the damping ratio $\zeta_i$ and the frequency of oscillation $\omega_i$ of the $i^{th}$ mode is given by $\zeta_i = -\sigma_i/\sqrt{\sigma_i^2 + \omega_i^2}$.

% Based on \eqref{eq:koopman_modes} and \eqref{eq:eigfun_def2}, the Koopman modes are given by: $\hat{L} = L^* B^\top$, where $w_i$ is the $i$-th left eigenvector of $K$ associated with eigenvalue $\mu_i$. The Koopman eigenfunctions are expressed as: $\Phi(x) = \Psi(x)\, \Xi$ where $\Xi$ contains the right eigenvectors of $K$, scaled such that $\hat{\ell}_i^* \xi_i = 1$. Unlike applications where EDMD is used mainly for signal reconstruction, this work applies it to identify critical oscillation modes and their contributors. Therefore, the $B$ matrix in \eqref{eq:B_def} is specified as a unity diagonal matrix.

Following \cite{Netto2019}, the data-driven participation factor for nonlinear systems is computed by combining both the left and right eigenvectors of $M_K$ using \eqref{eqn:righteigenvector} and \eqref{eqn:lefteigenvector}. These participation factors provide a quantitative measure of how each observable contributes to an oscillatory mode, thereby enabling identification of the most influential contributors to system oscillations. The participation factors $\mathcal{P}_i$ of the state variables in $i^{th}$ mode are defined as \cite{Netto2019}.

\begin{equation}
    \mathcal{P}_i =
    \begin{bmatrix}
        p_{1i} \\
        p_{2i} \\
        \vdots \\
        p_{ni}
    \end{bmatrix}
    =
    \begin{bmatrix}
        \hat{\phi}_{1i} \hat{\xi}_{i1} \\
        \hat{\phi}_{2i} \hat{\xi}_{i2} \\
        \vdots \\
        \hat{\phi}_{ni} \hat{\xi}_{in}
    \end{bmatrix},
    \label{eq:participation_factor}
\end{equation}
% \hlText{
% \begin{align}\nonumber
%     \mathcal{P}_i^{\top} &=
%     \begin{bmatrix}
%         p_{1i} & p_{2i} & \cdots & p_{ni}
%     \end{bmatrix}\\
%     &=
%     \begin{bmatrix}
%         \hat{\phi}_{1i} \hat{\xi}_{i1} & \hat{\phi}_{2i} \hat{\xi}_{i2} & \cdots & \hat{\phi}_{ni} \hat{\xi}_{in}
%     \end{bmatrix},
%     \label{eq:participation_factor}
% \end{align}
% }
%
where $p_{si}$ represents the relative participation of the $s^{th}$ state variable in the $i^{th}$ mode. A large participation factor value for a given state variable indicates that the variable significantly influences the corresponding mode. 

Since participation factors are scalars, the overall participation factor $\mathcal{P}_{ji}$ of a generation plant (say, $j^{th}$ plant) in $i^{th}$ poorly damped mode is the sum of the participation factors the observables associated with that plant  \cite{9705099}. 

% For a model-based approach, $\mathcal{P}_{ji}$ is obtained by summing the participation factors of all the state variables of an IBR while for data-driven methods (such as eDMD and DMD), $\mathcal{P}_{ji}$ is calculated as the sum of the participation factors of the observables (derived from measurements) at the point of connection (PoC) of the IBR with the network
%
\begin{equation}
   \mathcal{P}_{ji} = \sum_{s=1}^{N_{sj}} p_{si},  
    \label{eq:PF_def}
\end{equation}
where $p_{si}$ is the participation of state $s$ of plant $p$ in mode $i$ and $N_{sj}$ is the total number of states in plant $j$.

The normalized participation factor $\mathcal{P}_{ji}^\mathcal{N}$ of the $j^{th}$ plant in the $i^{th}$ poorly damped mode is given by
\begin{equation}
   \mathcal{P}_{ji}^\mathcal{N} = \frac{\mathcal{P}_{ji}}{\max_{ j\in [1,N]}\mathcal{P}_{ji}},   
    \label{eq:NPF_def}
\end{equation}
where $N$ is the number of generation plants with PMU. It should be noted that the proposed framework is not intended to replicate model-based participation factor results. Instead, it captures the system dynamics through the Koopman operator estimated via EDMD and introduces a data-driven participation factor, derived from the finite-dimensional subspace of the estimated Koopman operator, to identify the major contributors to poorly damped oscillations.
% We use the normalized participation factor $\mathcal{P}_{ji}^\mathcal{N}$ to compare the accuracy of the data-driven DMD and eDMD methods with respect to the actual model in the next section.

The proposed EDMD-based framework is applicable to both natural and forced oscillations. This is because EDMD approximates the Koopman operator within the subspace of chosen observables, capturing dynamics arising from both state variables and forcing signals (if any) embedded in the observables. By analyzing the resulting data-driven Koopman operator, the proposed method enables effective identification of major oscillation contributors under diverse operating conditions. 
% A rigorous mathematical proof of the applicability of the proposed framework to various oscillatory events is currently under consideration.

\subsection{Application Steps}
The steps to apply the proposed method are:

\begin{enumerate}

\item  Filter measurement noise and remove outliers or low-quality PMU samples to ensure reliable input data.

\item Identify dominant oscillatory modes, $f_s$, based on FFT of the active and reactive power time-series data.

\item Choose a data window covering at least 5 cycles of $f_s$.

\item Apply bandpass filter centered at $f_s$ with cutoff frequencies $f_c = [0.9f_s,\ 1.1f_s]$ and retain the central portion of the filtered data to mitigate boundary effects.

\item  Approximate the Koopman operator $M_K$ using \eqref{MK_def} with the given finite-dimensional observables: $P$ and $Q$ .

\item  Calculate the reduced order $\Tilde{M}_K$ using \eqref{eqn:hatKmatrix} with the reduced order $r$ determined based on the elbow point of the singular value spectrum.

\item Identify the poorly damped oscillation modes using~\eqref{eqn:eigvals} and \eqref{eq:continuous_eigenvalue}

% \item Obtain right eigen vector $\Tilde{\Phi}$ of $\Tilde{M}_K$ as ~\eqref{eqn:eigvals}

\item Calculate the right and left eigenvectors, $\hat{\Phi}$, $\hat{\Xi}$ by projecting back to the original higher-dimensional subspace using \eqref{eqn:righteigenvector} and \eqref{eqn:lefteigenvector}

\item  For each identified poorly damped mode $\lambda_i$, calculate data-driven participation factors using the corresponding elements of $\hat{\Phi}$, $\hat{\Xi}$ as per \eqref{eq:participation_factor}-\eqref{eq:NPF_def}

% \item  Perform data-driven modal analysis using \eqref{eq:continuous_eigenvalue}-\eqref{eq:NPF_def} to identify critical oscillation modes and compute participation factors of contributing components.

\end{enumerate}

\section{Case Study on Modified IEEE 39-bus System}
\begin{figure}[htbp]
    \centering
    \includegraphics[width=\columnwidth]{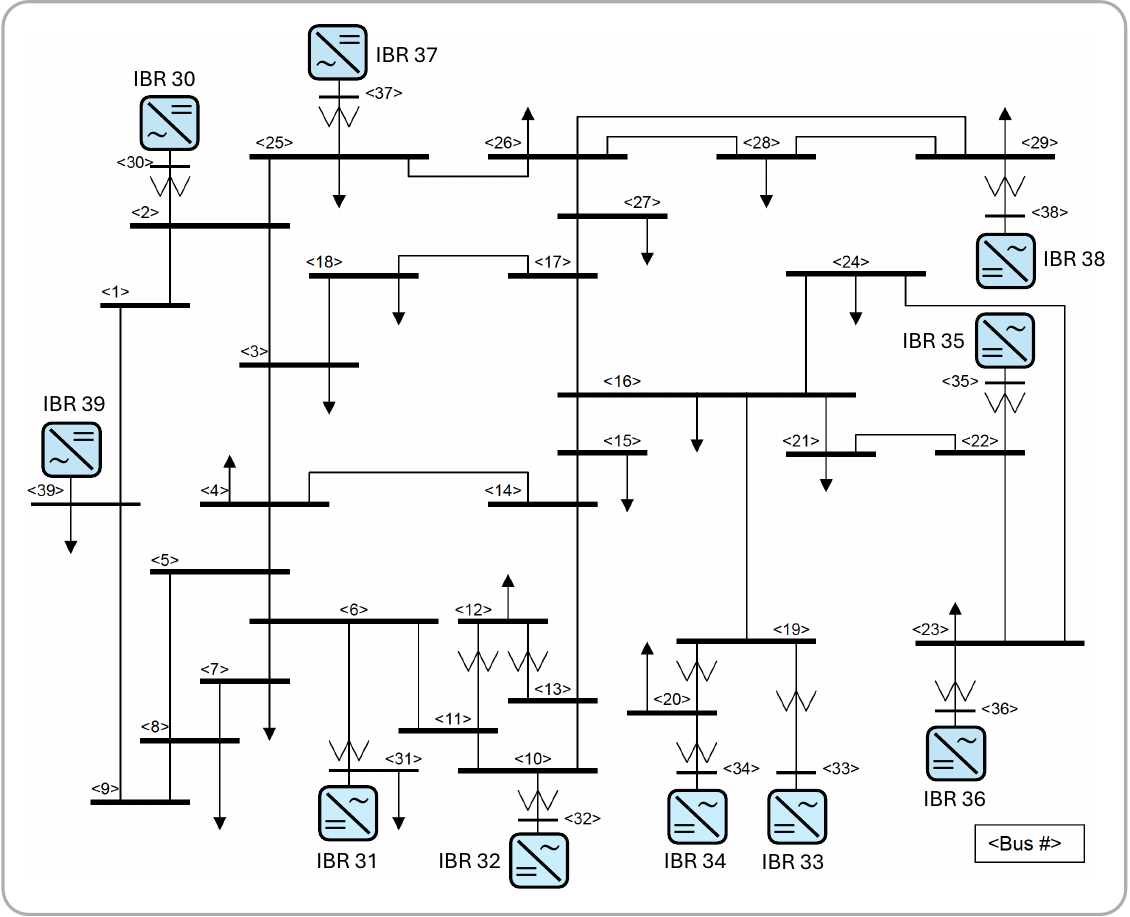} 
    \caption{Modified IEEE 39-bus test system with all synchronous machines replaced by IBRs. Different mix of GFL and GFM are considered as summarized in Table~\ref{tab:gfl_gfm}}
    \label{fig:network}
\end{figure}

First, we use a modified version of the IEEE 39-bus test system \cite{IEEE_benchmark_systems} to validate the proposed method. To mimic an IBR-dominated system, all synchronous machines in the original system are replaced by equivalent IBRs, as shown in Figure~\ref{fig:network}. Both grid-following (GFL) and grid-forming (GFM) IBRs are modeled to regulate active power $(P_r)$ and AC voltage magnitude $(V_r)$ at their respective point of interconnection (POI) \cite{Javaid2025}. 
% The control gains and filter parameters are provided in Appendix \ref{appendix_IBR}.

Four scenarios are considered to demonstrate the effectiveness of the proposed methodology in identifying major contributors to IBR-driven SSOs. The locations and active power outputs of GFLs and GFMs vary across scenarios, as summarized in Table~\ref{tab:gfl_gfm}, with each scenario designed to exhibit poorly damped oscillations. As representative examples, Scenarios 1 and 4 are examined in detail, and in each case, the major contributors identified from EDMD-based participation factors are compared against those obtained from actual model-based modal analysis.

% \begin{table}[h] 
% \centering 
% \caption{Scenarios with different mix of GFL and GFM and their active power $P$ output} \renewcommand{\arraystretch}{1.2} 
% \setlength{\tabcolsep}{9pt} 
% \begin{tabular}{|c|c|c|c|c|}

% \hline \textbf{Bus \#} & \multicolumn{2}{c|}{\textbf{Scenario \#1}} & \multicolumn{2}{c|}{\textbf{Scenario \#4}} \\ \hline & & $P$ (pu) & & $P$ (pu) \\ \hline \hline 30 & \cellcolor{orange!30}GFM & 2.75 & \cellcolor{orange!30}GFM & 2.5 \\ \hline 31 & \cellcolor{green!10}GFL & 4.99 & \cellcolor{orange!30}GFM & 12.35 \\ \hline 32 & \cellcolor{orange!30}GFM & 6.5 & \cellcolor{orange!30}GFM & 6.5 \\ \hline 33 & \cellcolor{green!10}GFL & 6.32 & \cellcolor{orange!30}GFM & 6.32 \\ \hline 34 & \cellcolor{orange!30}GFM & 5.08 & \cellcolor{orange!30}GFM & 5.08 \\ \hline 35 & \cellcolor{green!10}GFL & 6.5 & \cellcolor{green!10}GFL & 5.2 \\ \hline 36 & \cellcolor{orange!30}GFM & 5.6 & \cellcolor{green!10}GFL & 4.48 \\ \hline 37 & \cellcolor{green!10}GFL & 5.4 & \cellcolor{green!10}GFL & 4.32 \\ \hline 38 & \cellcolor{orange!30}GFM & 8.3 & \cellcolor{green!10}GFL & 6.64 \\ \hline 39 & \cellcolor{green!10}GFL & 10 & \cellcolor{green!10}GFL & 8 \\ \hline \hline GFL & \multicolumn{2}{c|}{54.1\%} & \multicolumn{2}{c|}{46.7\%} \\ \hline GFM & \multicolumn{2}{|c|}{45.9\%} & \multicolumn{2}{c|}{53.3\%} \\ \hline 
% \end{tabular} 
% \label{tab:gfl_gfm} 
% \end{table}

\begin{table}[ht!]
    \centering
    \caption{Four scenarios with different mix of GFL and GFM and their active power $P$ output}
    \renewcommand{\arraystretch}{1.2} % Increase row height for better readability
    \setlength{\tabcolsep}{3pt} % Reduce column spacing
    %\scriptsize % Shrink overall text size
    %\begin{tabular}{|c|p{0.8cm}|p{0.8cm}|p{0.8cm}|p{0.8cm}|p{0.8cm}|p{0.8cm}|p{0.8cm}|p{0.8cm}|}
    \begin{tabular}{|c|c|c|c|c|c|c|c|c|}
        \hline
        \textbf{Bus \#} & \multicolumn{2}{c|}{\textbf{Scenario \#1}} & \multicolumn{2}{c|}{\textbf{Scenario \#2}} & \multicolumn{2}{c|}{\textbf{Scenario \#3}} & \multicolumn{2}{c|}{\textbf{Scenario \#4}} \\
        \hline
        &  & $P$ (pu) &  & $P$ (pu) &  & $P$ (pu) &  & $P$ (pu) \\
        \hline \hline
        30 & \cellcolor{orange!30}GFM & 2.75 & \cellcolor{green!10}GFL & 3.55 & \cellcolor{green!10}GFL & 2.5 & \cellcolor{orange!30}GFM & 2.5 \\ \hline
        31 & \cellcolor{green!10}GFL & 4.99 & \cellcolor{orange!30}GFM & -5.86 & \cellcolor{green!10}GFL & 8.76 & \cellcolor{orange!30}GFM & 12.35 \\ \hline
        32 & \cellcolor{orange!30}GFM & 6.5 & \cellcolor{green!10}GFL & 9.23 & \cellcolor{green!10}GFL & 6.5 & \cellcolor{orange!30}GFM & 6.5 \\ \hline
        33 & \cellcolor{green!10}GFL & 6.32 & \cellcolor{orange!30}GFM & 6.32 & \cellcolor{green!10}GFL & 6.32 & \cellcolor{orange!30}GFM & 6.32 \\ \hline
        34 & \cellcolor{orange!30}GFM & 5.08 & \cellcolor{green!10}GFL & 7.21 & \cellcolor{green!10}GFL & 5.08 & \cellcolor{orange!30}GFM & 5.08 \\ \hline
        35 & \cellcolor{green!10}GFL & 6.5 & \cellcolor{orange!30}GFM & 6.5 & \cellcolor{orange!30}GFM & 5.85 & \cellcolor{green!10}GFL & 5.2 \\ \hline
        36 & \cellcolor{orange!30}GFM & 5.6 & \cellcolor{green!10}GFL & 7.95 & \cellcolor{orange!30}GFM & 5.04 & \cellcolor{green!10}GFL & 4.48 \\ \hline
        37 & \cellcolor{green!10}GFL & 5.4 & \cellcolor{orange!30}GFM & 5.4 & \cellcolor{orange!30}GFM & 4.86 & \cellcolor{green!10}GFL & 4.32 \\ \hline
        38 & \cellcolor{orange!30}GFM & 8.3 & \cellcolor{green!10}GFL & 11.79 & \cellcolor{orange!30}GFM & 7.47 & \cellcolor{green!10}GFL & 6.64 \\ \hline
        39 & \cellcolor{green!10}GFL & 10 & \cellcolor{orange!30}GFM & 10 & \cellcolor{orange!30}GFM & 9 & \cellcolor{green!10}GFL & 8 \\ \hline
        \hline 
        GFL & \multicolumn{2}{c|}{54.1\%} & \multicolumn{2}{c|}{64\%} & \multicolumn{2}{c|}{47.5\%} & \multicolumn{2}{c|}{46.7\%} \\
        \hline
        GFM &\multicolumn{2}{|c|}{45.9\%}  & \multicolumn{2}{c|}{36\%} & \multicolumn{2}{c|}{52.5\%} & \multicolumn{2}{c|}{53.3\%} \\

        \hline
    \end{tabular}
    \label{tab:gfl_gfm}
\end{table}

For each scenario, a 5\% step change in voltage reference at one of the ten IBRs is used to generate the data matrix $X, Y$ in~\eqref{eqn:dataXY} from the EMT simulation with a step size of 10 {\textmu}s. It should be noted that the step change is applied solely to excite SSO. The EDMD method does not require knowledge of the amplitude or the location of the step input. Simulated data are sampled at 30 Hz, which is a typical PMU reporting rate. Based on the discussion in Section \ref{subsec:Sel_of_Obser}, the active $P$ and reactive $Q$ power calculated from PMU measurement at POI of each IBR plant is used for EDMD. The data from 0.1 to 3 s time window is used to capture and analyse 5–15 Hz IBR-driven SSOs. Modal analysis of the linearized state-space model for each scenario is used as the benchmark to determine the accuracy of the data-driven participation factors $\mathcal{P^N}$ in \eqref{eq:NPF_def} obtained using EDMD. Table \ref{tab:IBRNDosc-source} summarizes the damping ratios $\zeta$ (in \%) and frequencies $\omega$ (in Hz) of poorly damped modes, together with their major contributors as identified by model-based (actual) and data-driven (estimated) approaches.

% Define a new column type for centered p-width columns
\newcolumntype{C}[1]{>{\centering\arraybackslash}p{#1}}

% Then in the body:
\begin{table}[ht!]
  \centering
  \caption{Natural oscillation cases in IEEE-39 bus test system}
  \label{tab:IBRNDosc-source}
  \setlength{\tabcolsep}{4pt}
  \renewcommand{\arraystretch}{1.3}
  \begin{tabular}{|c|C{1.1cm}|C{1.1cm}|C{1.1cm}|C{1.1cm}|C{1.1cm}|}
    \hline 
    Test case & Damping (\%) & \multicolumn{2}{c|}{Freq (Hz)} & \multicolumn{2}{c|}{Top contributors IBR\#} \\
    \cline{3-6}
     & & Actual & Estimated & Actual & Estimated \\
    \hline \hline

    Scenario\#1 & 0.78 & 9.34 & 9.39 & 30 & 30 \\
    \hline
    Scenario\#2 & 4.50 & 5.28 & 5.23 & 38 & \makecell{38} \\
    \hline
    Scenario\#3 & 2.84 & 10.98 & 10.96 & \makecell{35,36} & \makecell{35,36} \\
    \hline
    Scenario\#4 & \makecell{5.60 \\ 0.11} & \makecell{10.47 \\ 9.17} & \makecell{10.44 \\ 9.18} & \makecell{33,34\\30} & \makecell{33,34\\30} \\
    \hline
  \end{tabular}
\end{table}

\subsection{Scenario \#1}

\begin{figure}[htbp]
    \centering
    \includegraphics[width=\columnwidth]{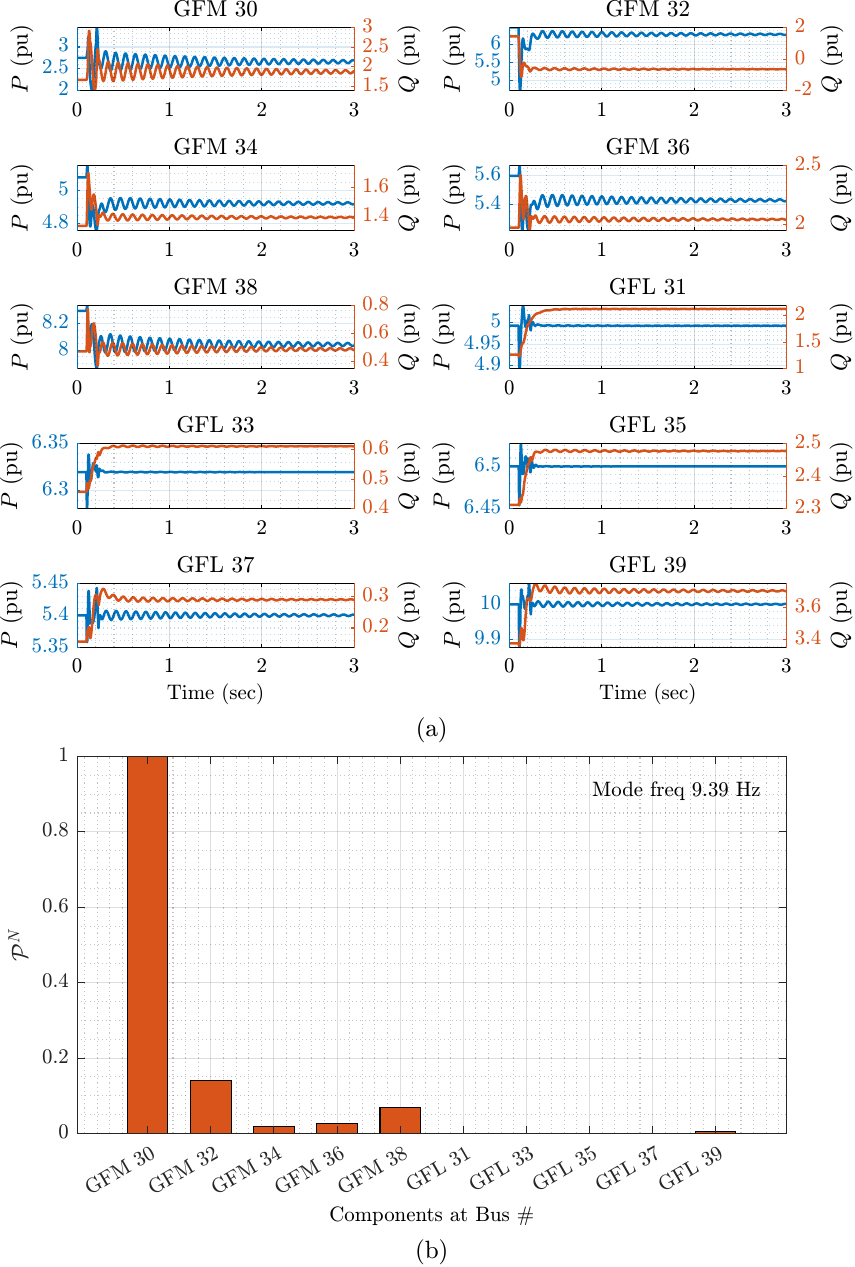} % Replace with the actual image filename
    \caption{EDMD-based analysis for Scenario \# 1: (a) time-series $P$ and $Q$ at POI of each IBR plant, (b) data-driven participation factor of each IBR plant}
    \label{fig:system_scenario1}
\end{figure}

The simulated $P$ and $Q$ at the POI of each GFL/GFM plant for scenario \#1, shown in Figure \ref{fig:system_scenario1}(a), are used to perform data-driven root-cause analysis using EDMD. FFT reveals a dominant oscillation at 9.34 Hz. Based on the singular value analysis, the truncated order $r$ is set to 2.  Table \ref{tab:IBRNDosc-source} shows that the frequency of the poorly damped SSO mode estimated by EDMD is closely aligned with actual model-based values. The resulting data-driven participation factors, shown in Figure \ref{fig:system_scenario1}(b), indicate that the GFM at bus 30 is the major contributor to 9.3 Hz oscillation, consistent with the model-based results. This demonstrates the effectiveness of the proposed EDMD-based approach in accurately pinpointing the dominant IBRs that contribute to SSOs.

% \begin{table}[htbp]
%     \centering
%     \caption{Damping and frequency of poorly damped SSO for Scenario \#1 }
%     \label{tab:eigenvalue_comparison1}
%     %\setlength{\tabcolsep}{6pt} % Adjust column spacing
%     \renewcommand{\arraystretch}{1.2} % Adjust row height
%     \small % Ensure font size matches surrounding text
%     \begin{tabular}{|l|c|c|}
%         \hline
%         \textbf{Method} & $\omega$ (Hz) & $\zeta$ (\%) \\
%         \hline \hline
%         Model based &  9.33 & 1.43 \\
%         \hline
%         Data-driven eDMD  &  9.34 & 1.20 \\
%         \hline
%         Data-driven DMD   &  9.33 & 1.23 \\
%         \hline
%     \end{tabular}
% \end{table}

% \begin{figure}[htbp]
%     \centering
%     \includegraphics[width=\columnwidth]{Figure/PF_plots_scenario1.png} 
%     \caption{Data-driven participation factors from eDMD and DMD with respect to actual model-based for Scenario \#1}
%     \label{fig:pf1}
% \end{figure}
\subsection{Scenario \#4}

\begin{figure}[htbp]
    \centering
    \includegraphics[width=\columnwidth]{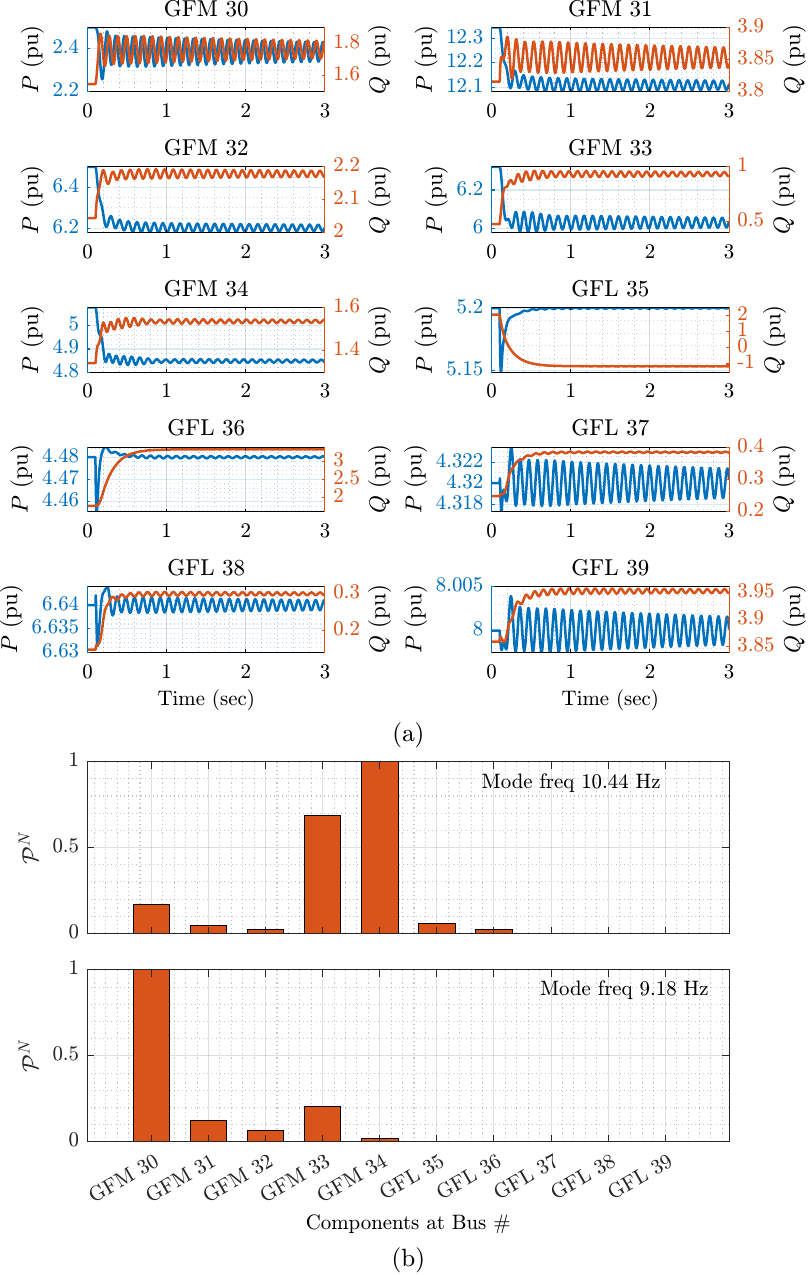} % Replace with the actual image filename
    \caption{EDMD-based analysis for Scenario \# 4: (a) time-series $P$ and $Q$ at POI of each IBR plant, (b) data-driven participation factor of each IBR plant.}
    \label{fig:system_scenario4}
\end{figure}

Figure \ref{fig:system_scenario4}(a) shows the simulated P and Q time-series data for Scenario \#4, which are used for data-driven modal analysis using EDMD. Unlike Scenario \#1, the FFT spectrum here contains two peaks, indicating poorly damped modes at 10.47 Hz and 9.17 Hz. Based on the singular value analysis, truncated orders of $r=$ 5 and 2 are selected for these modes, respectively. Table \ref{tab:IBRNDosc-source} shows that the estimated frequencies of the poorly damped SSO modes obtained using EDMD closely align with the actual model-based values. Figure \ref{fig:system_scenario4}(b) illustrates data-driven participation factors obtained from EDMD. Both EDMD and actual model-based analysis identify GFMs at bus 33 and bus 34 as the main contributors in the 10.46 Hz SSO mode. Similarly, EDMD correctly determines that the GFM at bus 30 has the highest participation in the 9.16Hz SSO mode.

% \begin{figure}[htbp]
%     \centering
%     \includegraphics[width=\columnwidth]{Figure/PF_plots_scenario4.png} 
%     \caption{Data-driven participation factors from eDMD and DMD with respect to actual model-based for Scenario \#4}
%     \label{fig:pf4}
% \end{figure}

This scenario highlights that the amplitude of the oscillation at a particular IBR is not necessarily a correct indication of its participation in a poorly damped SSO mode. For instance, Figure \ref{fig:system_scenario4}(a) shows relatively small oscillation amplitudes for both $P$ and $Q$ at the GFMs at buses 33 and 34. However, these two GFMs have the highest participation factor (as obtained from the model, EDMD) in the 10.4 Hz SSO mode. This finding highlights that identifying the relative contribution of IBRs based solely on oscillation magnitude at their POI could be misleading and not necessarily reliable.

\section{Case Study on WECC 179-bus Test System}

\begin{figure}[htbp]
    \centering
    \includegraphics[width=\columnwidth]{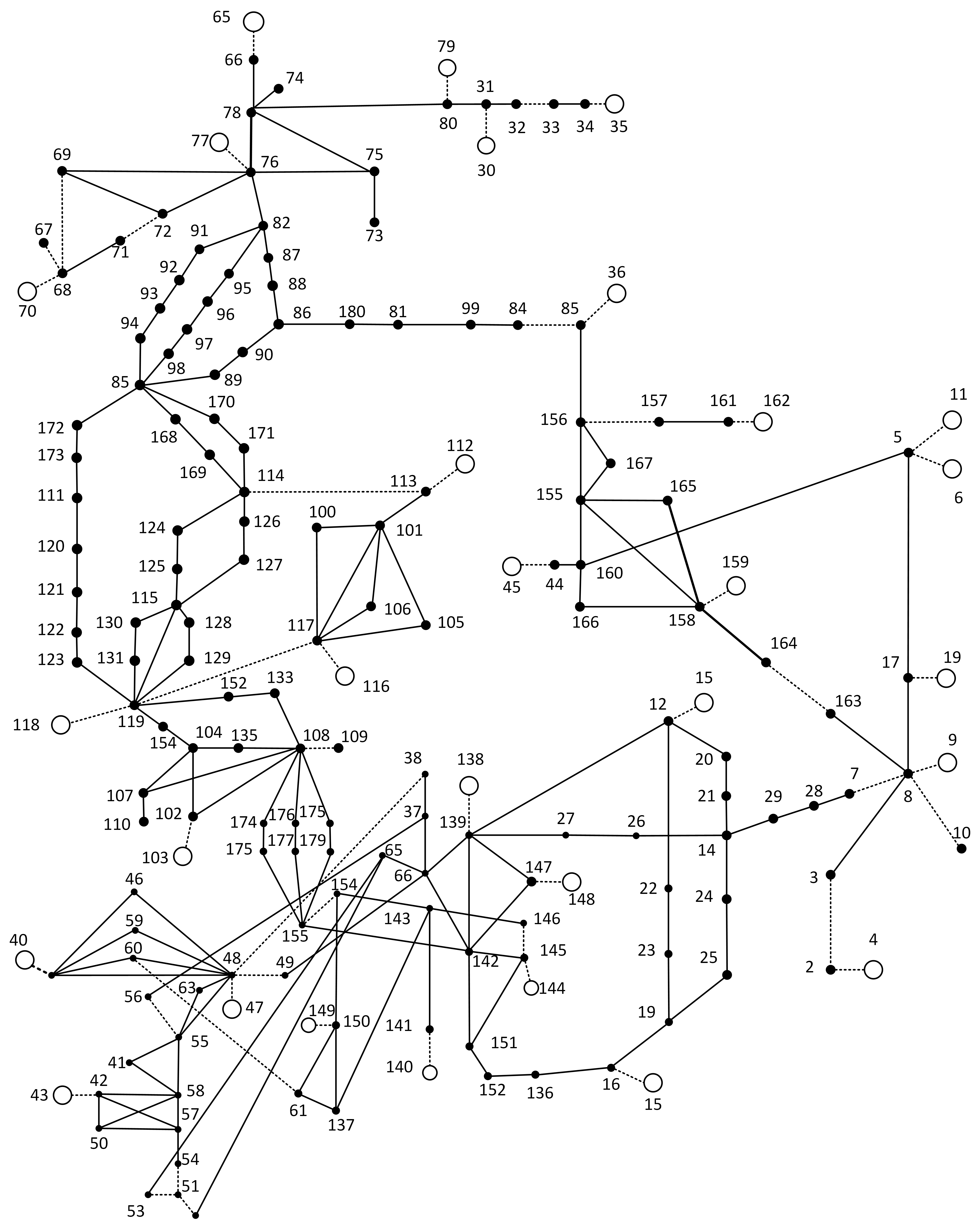}
    \caption{The scheme diagram of WECC 179-bus system. \cite{Maslennikov2016}}
    \label{fig:WECCsystem}
\end{figure}

To verify the effectiveness of the proposed data-driven method on larger test systems, the simulated cases on the WECC 179-bus system, from the online Power System Sustained Oscillation Library \cite{Maslennikov2016} are used. These cases demonstrate the method's ability to identify significant contributors to poorly damped electro-mechanical oscillations in SM-dominated networks. The diagram of WECC-179 bus system is shown in Figure \ref{fig:WECCsystem}.

% It is important to distinguish between forced and natural oscillations. Forced oscillations arise from continuous external forcing at the oscillatory frequency. Natural oscillations, on the contrary, are driven by energy exchanges among power generation plants. They emerge from the coupled system dynamics rather than from a single external source. Consequently, this study focuses on identifying the significant contributors to the poorly damped oscillation to support effective mitigation.

The proposed EDMD-based framework is applied to analyze low-frequency oscillations under various cases. Due to space limitations, only four test cases - two natural and two forced oscillations - are presented here. For WECC 179-bus cases, a measurement window from 5 to 40 s, where oscillations are prominant, is used to capture the 0.2–2 Hz low-frequency SM-driven oscillation dynamics. The detailed analysis of each test case is presented in the following subsections to validate the proposed EDMD-based root-cause analysis framework. 

% All analyses are conducted using MATLAB R2023b.

% Subsequently, the method is further validated using actual oscillatory events from real PMU data in the ISO-NE network.

\subsection{Natural Oscillation}

For WECC 179 bus test cases, the natural oscillations are caused by a detuned SM with reduced damping/negative factor. These oscillations are triggered by faults. The test cases are summarized in Table~\ref{tab:NDosc-source}.

\newcolumntype{C}[1]{>{\centering\arraybackslash}p{#1}}
\begin{table}[ht!]
  \caption{Natural oscillation cases in WECC-179 bus test system \cite{Maslennikov2016}}
  \label{tab:NDosc-source}
  \centering
  \setlength{\tabcolsep}{2pt}
  \renewcommand{\arraystretch}{1.5}
  \begin{tabular}{|c|c|c|C{1.1cm}|C{1.1cm}|C{1.1cm}|C{1.1cm}|}
    \hline 
    Test case & Fault at & Damping  & \multicolumn{2}{c|}{Freq (Hz)} & \multicolumn{2}{c|}{ Top contributors SM\#} \\
    \cline{4-7}
     & bus\# & (\%) & Actual & Estimated & Actual & Estimated \\
    \hline \hline
    % Sim1ND & 159 & 0.01 & 1.41 & 1.40 & 45, 159 & 45, 159 \\
    % \hline
    % Sim2ND & 79  & 0.02 & 0.37 & 0.37 & 36, 65 & 36, 65 \\
    % \hline
    Sim6ND & 159 & -0.93 & 1.41 & 1.41 & 45, 159 & 45, 159 \\
    \hline
    Sim8ND & 159 & \makecell{1.06 \\ 0.22} & \makecell{1.27 \\ 1.41} & \makecell{1.26 \\ 1.39} & \makecell{45, 159,\\ 36} & \makecell{45, 159,\\ 36, 77} \\
    \hline
  \end{tabular}
\end{table}

\begin{figure}[htbp]
  \centering
    \includegraphics[width=\columnwidth]{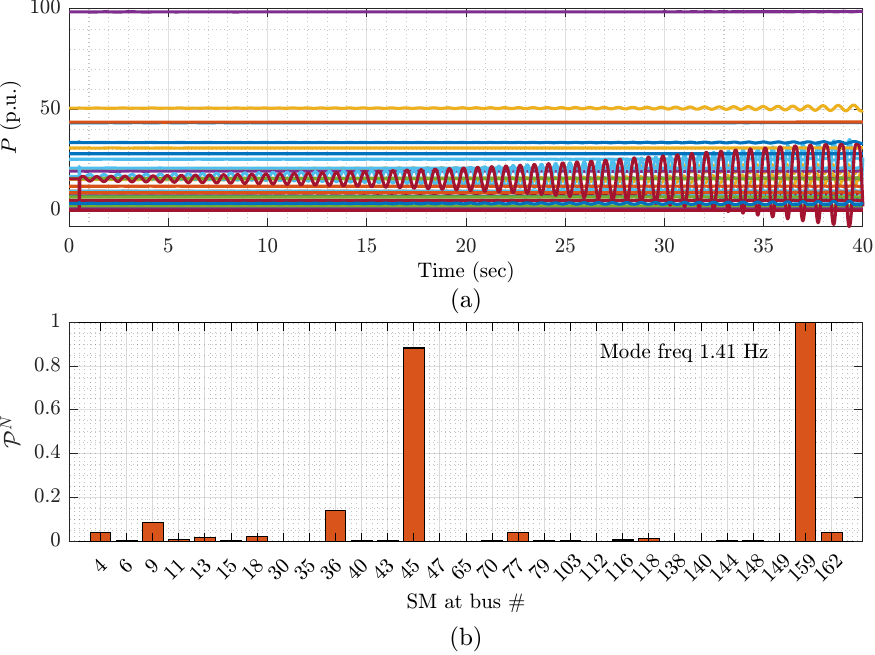}%
  \caption{EDMD-based analysis for Sim6ND: (a) time-series $P$ and $Q$ at POI of each SM generation plant, (b) data-driven participation factor of each SM.}
  \label{fig:sim6nd}
\end{figure}

For the first test case, Sim6ND, two SMs contribute significantly to the natural oscillation. Figure~\ref{fig:sim6nd} illustrates the data-driven oscillation contributors for this case. The simulated active power ($P$) data at the POI of each SM generation plant are shown in Figure~\ref{fig:sim6nd}(a). FFT of the time-domain data indicates a dominant mode at 1.41 Hz. The EDMD truncated order, $r$ is set to 6 from singular value analysis. The estimated frequency of the poorly damped mode obtained using the proposed EDMD method, as shown in Table \ref{tab:NDosc-source}, is the same as the actual model-based values for EDMD. Figure~\ref{fig:sim6nd}(b) presents the data-driven participation factors derived from EDMD, which clearly highlight SMs 45 and 159 as the dominant contributors to the 1.41 Hz oscillation. This result aligns closely with the model-based analytical findings in \cite{Maslennikov2016}. 

% \begin{figure}[!t]
%   \centering
%   \subfloat[Time-domain simulation]{%
%     \includegraphics[width=\columnwidth]{8ND_P_SG.png}%
%     \label{fig:sim8nd_p}}\\[1ex]
%   \subfloat[FFT analysis]{%
%     \includegraphics[width=\columnwidth]{8ND_FFT.png}%
%     \label{fig:sim8nd_fft}}\\[1ex]
%   \subfloat[Data-driven participation factors]{%
%     \includegraphics[width=\columnwidth]{8ND_PF.png}%
%     \label{fig:sim8nd_pf}}
%   \caption{Sim8ND results.}
%   \label{fig:sim8nd}
% \end{figure}

\begin{figure}[htbp]
  \centering
    \includegraphics[width=\columnwidth]{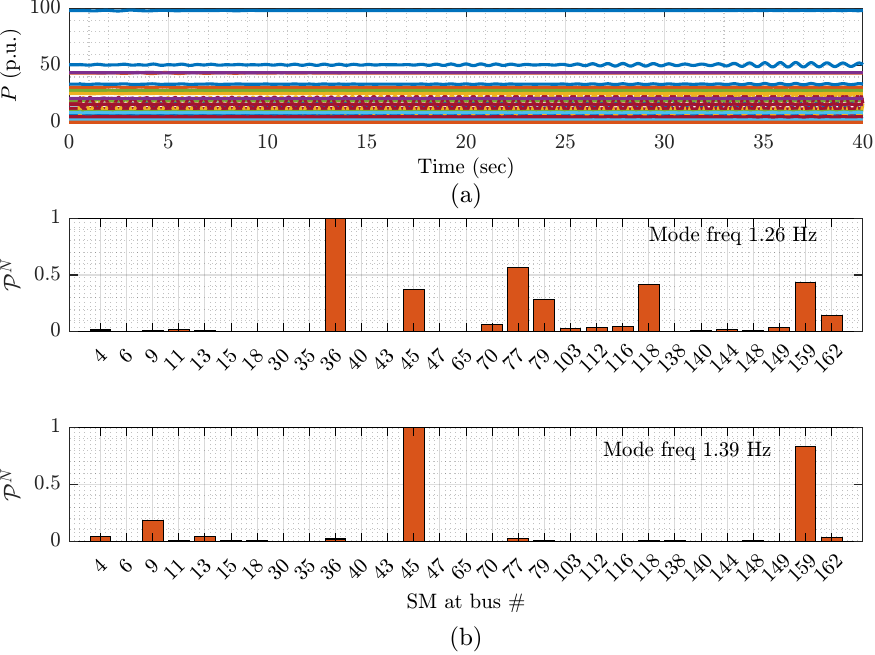}%
  \caption{EDMD-based analysis for Sim8ND: (a) time-series $P$ and $Q$ at POI of each SM generation plant, (b) data-driven participation factor of each SM.}
  \label{fig:sim8nd}
\end{figure}

The test case for the SimND8 case involves two poorly damped modes, is shown in Figure~\ref{fig:sim8nd}. 
FFT reveals two spectral peaks at 1.41 Hz and 1.27 Hz, indicating the presence of these two oscillatory modes. Based on the singular value analysis, truncated orders $r$ are set 4 and 7 for 1.41 Hz and 1.27 Hz modes, respectively. The estimated frequencies from EDMD closely match the actual poorly damped modes listed in Table \ref{tab:NDosc-source}, with errors of 0.79\% and 1.42\% for the 1.27 Hz and 1.41 Hz oscillations respectively. Both the participation factors derived from EDMD and those from the model-based analysis identify the dominant contributors as follows: SM 36 for the 1.26 Hz mode, and SMs 159 and 45 for the 1.39 Hz mode. These results are consistent with the setup in \cite{Maslennikov2016}, where negative damping is applied to SMs 36 and 159. 
% \hlText{However, EDMD also identifies sizable participation for SM 77, which is not far from SM 36 with strong contribution to the oscillations. This discrepancy is attributed to the absence of an optimally selected set of lifted observables for capturing system dynamics, highlighting the need for further research on lifted observable selection.}

\subsection{Forced Oscillation}

The data-driven root-cause analysis method is next tested for the forced oscillations category. Unlike natural oscillation events, a forced oscillation is sustained by a continuous periodic signal injected into the system, compelling it to oscillate at a specific frequency. Hence, there is a source where the forcing signal is injected. In this study, the forcing signal is applied to the excitation system of a specific SM to generate forced oscillations. As discussed in Section \ref{sec:Methodology}, this subsection demonstrates the effectiveness of the proposed EDMD-based data-driven method in detecting and performing root-cause analysis to identify the major contributors of forced oscillations. Table~\ref{tab:FOsc-source} compares the results of the data-driven analysis with the test case descriptions in \cite{Maslennikov2016} for the forced oscillation cases.

\newcolumntype{C}[1]{>{\centering\arraybackslash}p{#1}}
\begin{table}[ht!]
  \caption{Forced oscillation cases in WECC-179 bus test system \cite{Maslennikov2016}}
  \label{tab:FOsc-source}
  \centering
  \setlength{\tabcolsep}{4pt}
  \renewcommand{\arraystretch}{1.3}
  \begin{tabular}{|c|C{1.1cm}|C{1.1cm}|c|c|}
    \hline
    \multirow{2}{*}{Test case} & 
    \multicolumn{2}{c|}{Freq (Hz)} & 
    {Actual source} & 
    {Estimated top } \\
    \cline{2-3} 
     & Actual & Estimated & SM\# & contributors SM\# \\
    \hline \hline
    Sim1F   & 0.86 & 0.86 & 4 & \makecell{4} \\
    \hline
    Sim6F3  & 0.40 & 0.41 & 79 & \makecell{79} \\
    \hline
    % Sim7F1  & \makecell{0.65, 0.43} & \makecell{0.65, 0.43} & \makecell{79, 118} & \makecell{79, 118, 30, 77} \\
    % \hline
    % Sim7FM1 & \makecell{0.65, 0.43} & \makecell{0.65, 0.43} & \makecell{79, 118} & \makecell{79, 118, 77} \\
    % \hline
  \end{tabular}
\end{table}

% \begin{figure}[!t]
%   \centering
%   \subfloat[Time-domain simulation]{%
%     \includegraphics[width=\columnwidth]{1F_P_SG.png}%
%     \label{fig:sim1f_p}}\\[1ex]
%   \subfloat[FFT analysis]{%
%     \includegraphics[width=\columnwidth]{1F_FFT.png}%
%     \label{fig:sim1f_fft}}\\[1ex]
%   \subfloat[Data-driven participation factors]{%
%     \includegraphics[width=\columnwidth]{1F_PF.png}%
%     \label{fig:sim1f_pf}}
%   \caption{Sim1F results.}
%   \label{fig:sim1f}
% \end{figure}

\begin{figure}[htbp]
  \centering
    \includegraphics[width=\columnwidth]{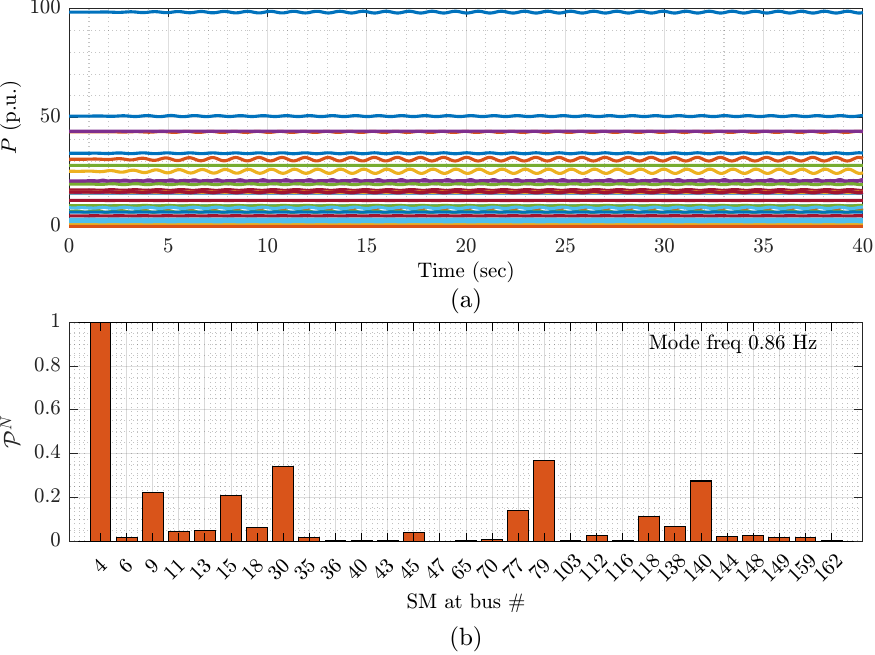}\

  \caption{EDMD-based analysis for Sim1F: (a) time-series $P$ and $Q$ at POI of each SM generation plant, (b) data-driven participation factor of each SM.}
  \label{fig:sim1f}
\end{figure}

The first case, Sim1F, involves a sinusoidal forcing signal injected into the excitation system of an SM, as shown in Figure~\ref{fig:sim1f}. FFT identifies a dominant frequency of 0.86 Hz, indicated by a clear peak in the spectrum. For this case, the EDMD truncated system order $r$ is set to 4 based on the singular value analysis. The estimated frequency from EDMD, as reported in Table \ref{tab:FOsc-source}, matches the actual injected forcing frequency. Data-driven participation factors in Figure~\ref{fig:sim1f}(b) identify SM 4 as the primary contributor with the highest participation in the 0.86 Hz mode, consistent with that in \cite{Maslennikov2016}. As the forcing signal resonates with the local mode and propagates through the network, other SMs, including SM 9, 15, 30, and 79 exhibit sizable contributions alongside SM 4.

% As a forced oscillation propagates through the network, nearby SMs, including SMs 9, and 15, are also driven to oscillate at the same frequency and therefore exhibit high participation values. However, based on the EDMD participation factor analysis, SM 4 is clearly identified as the source of the forcing signal, while the other generators’ high participation is attributed to their electrical proximity to the source.

% \begin{figure}[!t]
%   \centering
%   \subfloat[Time-domain simulation]{%
%     \includegraphics[width=\columnwidth]{6F3_P_SG.png}%
%     \label{fig:sim6f3_p}}\\[1ex]
%   \subfloat[FFT analysis]{%
%     \includegraphics[width=\columnwidth]{6F3_FFT.png}%
%     \label{fig:sim6f3_fft}}\\[1ex]
%   \subfloat[Data-driven participation factors]{%
%     \includegraphics[width=\columnwidth]{6F3_PF.png}%
%     \label{fig:sim6f3_pf}}
%   \caption{Sim6F3 results.}
%   \label{fig:sim6f3}
% \end{figure}

\begin{figure}[htbp]
  \centering
    \includegraphics[width=\columnwidth]{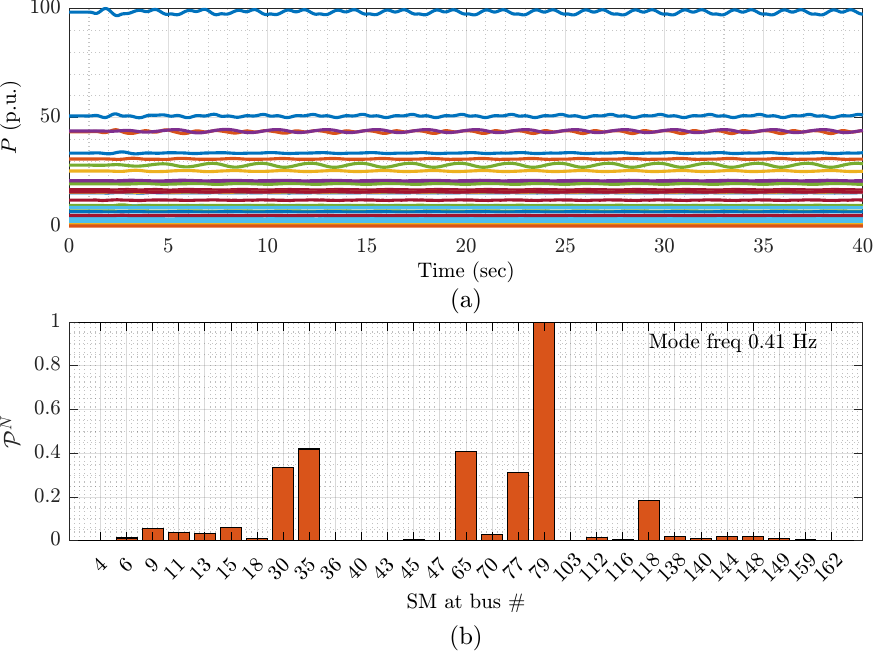}%
  \caption{EDMD-based analysis for Sim6F3: (a) time-series $P$ and $Q$ at POI of each SM generation plant, (b) data-driven participation factor of each SM.}
  \label{fig:sim6f3}
\end{figure}

For test case Sim 6F3, the forcing signal is rectangular rather than sinusoidal. The rectangular forcing signal contains odd-order harmonics of the fundamental oscillating frequency. FFT confirms the presence of these harmonics in the system. The dominant frequencies of the conjugate pairs include 1.2 Hz, 0.8 Hz, and 0.4 Hz, all integer multiples of 0.4 Hz. Accordingly, 0.4 Hz is determined as the dominant frequency. The EDMD truncated order $r$ is set to 5 based on the singular value analysis. The EDMD-estimated forced oscillation frequency is 0.41 Hz, closely matching the actual forcing frequency. The data-driven participation factor shown in Figure \ref{fig:sim6f3}(b) correctly identifies SM 79 where the forcing signal is injected, as the source of the 0.4 Hz oscillation. The forced oscillation also propagates through the network, resulting in sizable participation from nearby SMs 30, 35, 65, and 77.
% , while also indicating high participation from nearby SMs 77, and 30.

\section{Case Study with Real PMU Data from ISO-New England}
\begin{figure}[htbp]
    \centering
    \includegraphics[width=\columnwidth]{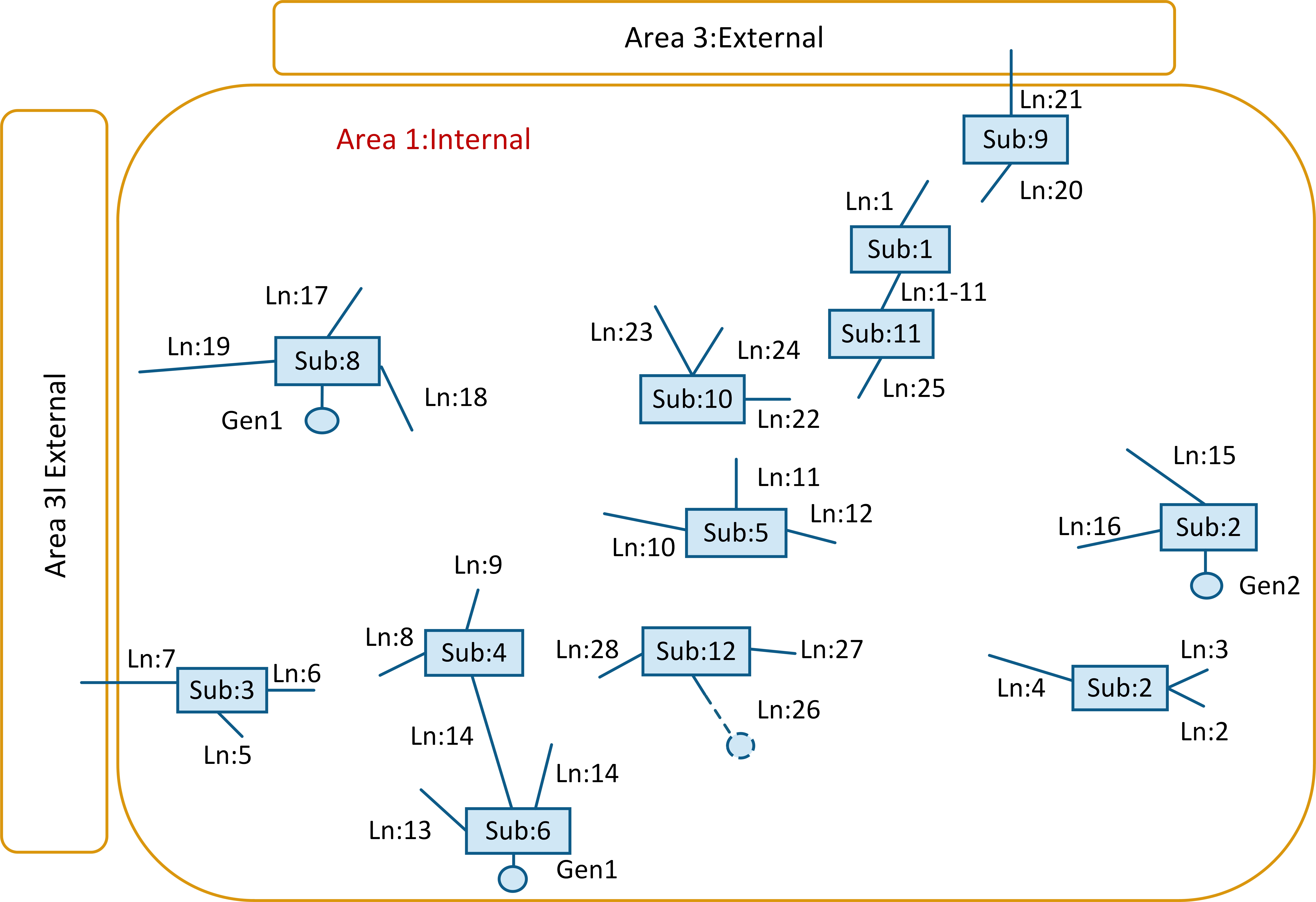}
    \caption{The ISO New England map.\cite{Maslennikov2016}}
    \label{fig:ISOsystem}
\end{figure}

To validate the effectiveness of the proposed data-driven method for real oscillatory events using PMU measurements captured during actual oscillatory events in the ISO-NE system \cite{Maslennikov2016} are analysed. These events occurred in ISO New England system shown in Figure \ref{fig:ISOsystem}. Similar to the simulated cases, this section is divided into natural oscillation and forced oscillation categories. However, unlike simulated cases, real-world measurements often contain noise, outliers, and poor-quality samples. 

% Moreover, practical power systems rarely have extensive PMU coverage, resulting in only partial system observability.

Accordingly, appropriate preprocessing is essential before applying the data-driven method. Measurements with poor quality and outlier values are removed, and a low-pass filter (LPF) is applied to suppress measurement noise. Given the low-frequency range of oscillations of interest (0.1 Hz-2 Hz), a 3 Hz cut-off frequency is used for the LPF in this study.

The proposed EDMD-based framework is applied to several of real power system oscillatory events using the archived PMU measurements. Due to space limitations, two representative cases, Case 1 (forced oscillation) and Case 3(natural oscillation), are presented in this paper. Table \ref{tab:ISOOsc-source} summarizes the event information and the corresponding results obtained from the EDMD-based analysis.

\newcolumntype{C}[1]{>{\centering\arraybackslash}p{#1}}
\begin{table}[ht!]
  \caption{Real Oscillatory Event in ISO-NE}
  \label{tab:ISOOsc-source}
  \centering
  \setlength{\tabcolsep}{4pt}
  \renewcommand{\arraystretch}{1.3}
  \begin{tabular}{|c|C{1.1cm}|C{1.1cm}|c|c|}
    \hline
    \multirow{2}{*}{Case\#} & 
    \multicolumn{2}{c|}{Freq (Hz)} & 
    {Actual source} & 
    {Estimated top } \\
    \cline{2-3} 
     & Actual & Estimated &  & contributors  \\
    \hline \hline
    1   & 0.27 & 0.28 & \makecell{Generator at \\Area 2} & \makecell{Sub3:Ln7 \\(Generator at \\Area 2)} \\
    \hline
    3  & 1.13 & 1.15 & \makecell{Sub2:Ln2, Ln4} & \makecell{Sub2:Ln2, Ln4} \\
    \hline

  \end{tabular}
\end{table}

\subsection{ISO-NE Case 3 – Natural Oscillation}
\begin{figure}[htbp]
  \centering
    \includegraphics[width=\columnwidth]{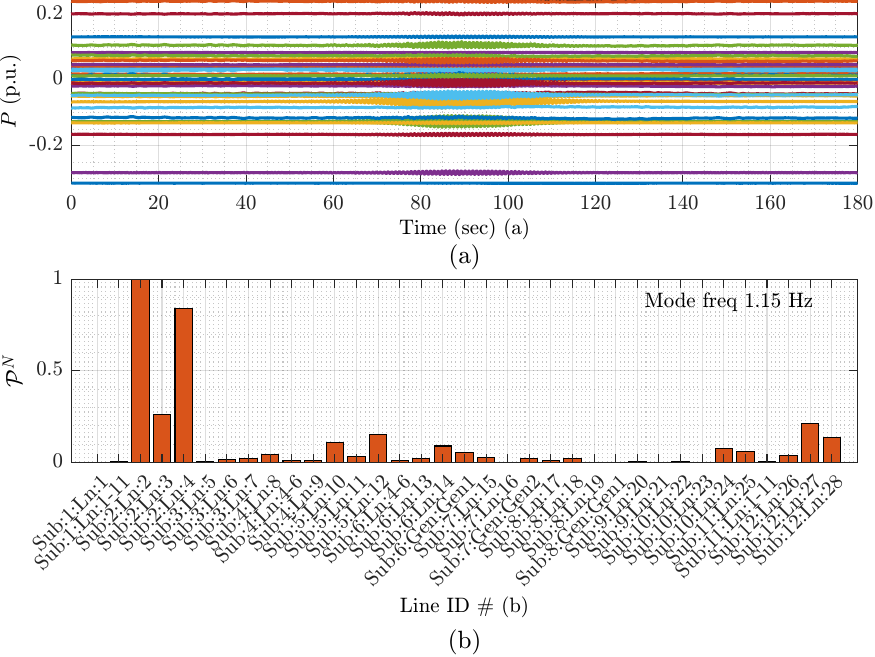}%
  \caption{EDMD-based analysis for ISO\_NEcase3: (a) time-series $P$ and $Q$ from each monitored PMU, (b) data-driven participation factors.}
  \label{fig:iso_necase3}
\end{figure}

% \begin{figure}[!t]
%   \centering
%   \subfloat[Time-domain simulation]{%
%     \includegraphics[width=\columnwidth]{ISO_NEcase3_P_SG.png}%
%     \label{fig:iso_necase3_p}}\\[1ex]
%   \subfloat[FFT analysis]{%
%     \includegraphics[width=\columnwidth]{ISO_NEcase3_FFT.png}%
%     \label{fig:iso_necase3_fft}}\\[1ex]
%   \subfloat[Data-driven participation factors]{%
%     \includegraphics[width=\columnwidth]{ISO_NEcase3_PF.png}%
%     \label{fig:iso_necase3_pf}}
%   \caption{ISO\_NEcase3 results.}
%   \label{fig:iso_necase3}
% \end{figure}
The data-driven analysis for ISO-NE Case 3 is presented in Figure~\ref{fig:iso_necase3}. The pre-processed time-series data are shown in Figure~\ref{fig:iso_necase3}(a). FFT of the filtered time-series data reveals a prominent spectral peak at 1.13 Hz. To capture the system dynamics, the oscillatory portion of the time-series data, between 70 s to 120 s, is used as input to the EDMD analysis. The truncated order $r$ is set to 3 as per the singular value analysis. 

The proposed method correctly identifies the dominant oscillation at 1.15 Hz, which is closed to the actual oscillation frequency in Table \ref{tab:ISOOsc-source}. The resulting EDMD-based participation factors point to devices connected to Lines 2 and 4 at the Substation 2 as the main contributors to the oscillation, closely matching the actual source reported in \cite{Maslennikov2016}. A smaller contribution is observed from Line 3 connected to Substation 2 compared with Line 2 and 4, suggesting that the oscillation originates in the eastern part of the system and propagates westward.

\subsection{ISO-NE Case 1 – Forced Oscillation}
\begin{figure}[htbp]
  \centering
    \includegraphics[width=\columnwidth]{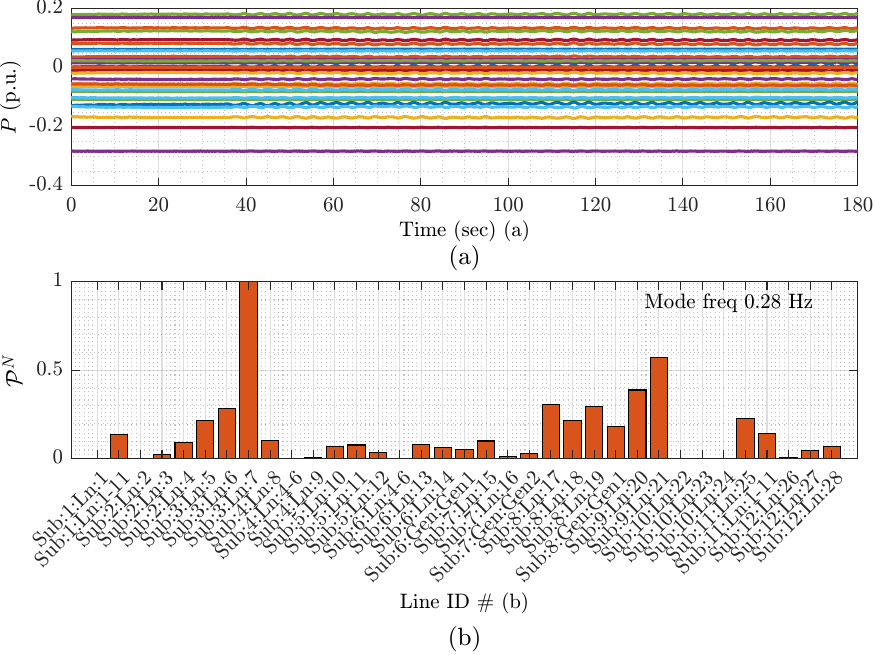}%
  \caption{EDMD-based analysis for ISO\_NEcase1: (a) time-series $P$ and $Q$ from each monitored PMU, (b) data-driven participation factors.}
  \label{fig:iso_necase1}
\end{figure}

% \begin{figure}[!t]
%   \centering
%   \subfloat[Time-domain simulation]{%
%     \includegraphics[width=\columnwidth]{ISO_NEcase1_P_SG.png}%
%     \label{fig:iso_necase1_p}}\\[1ex]
%   \subfloat[FFT analysis]{%
%     \includegraphics[width=\columnwidth]{ISO_NEcase1_FFT.png}%
%     \label{fig:iso_necase1_fft}}\\[1ex]
%   \subfloat[Data-driven participation factors]{%
%     \includegraphics[width=\columnwidth]{ISO_NEcase1_PF.png}%
%     \label{fig:iso_necase1_pf}}
%   \caption{ISO\_NEcase1 results.}
%   \label{fig:iso_necase1}
% \end{figure}
Next, the proposed framework is tested in ISO-NE Case 1, for a forced oscillation. Here, the oscillation source is located outside the monitored area. The preprocessed PMU time series for this case is shown in Figure~\ref{fig:iso_necase1}(a). FFT reveals a single dominant spectral peak at 0.27 Hz. The oscillatory transient from 40 s to 180 s is fed for EDMD analysis, and based on the singular value analysis of the filtered data, the truncated order is set to 3.

The EDMD-based analysis results, shown in Figure~\ref{fig:iso_necase1}(b), identify a poorly damped mode at 0.28 Hz, close to the actual oscillation frequency. The data-driven participation factors highlight Lines 7 connected to Substation 3 having significant contributions in the oscillation. In particular, Line 7 (Substation 3) exhibits notably higher participation values compared to other lines and generators at the same substations. This indicates that the oscillations propagate from Area 2 toward the east, consistent with the source location reported in \cite{Maslennikov2016}.

% \section{Discussion}
% The sustained oscillatory event library for the WECC system and the real oscillatory events in ISO-NE have previously been analyzed for root-cause identification using methods such as dissipated energy flow \cite{Maslennikov2017} and DMDc-based analysis \cite{10328697}. Compared with these approaches, the proposed EDMD-based method offers several advantages. It not only identifies the oscillatory frequency but also quantifies the contribution of individual components to the oscillation, rather than attributing it to a single source. This feature makes the method particularly suitable for analyzing IBR-driven natural oscillations, where instability arises from the dynamic interaction between IBRs and the network.

% By incorporating FFT and data filtering, the EDMD framework can also handle scenarios with multiple oscillatory modes. In such cases, it enables accurate root-cause analysis of each mode individually. Furthermore, unlike methods that require more extensive wide-area PMU coverage, the proposed approach only relies on measurements from the terminals of critical power components. Even with reduced PMU availability, the EDMD-based analysis is capable of identifying the key contributors to oscillations with high accuracy.

\section{Conclusion}
This paper demonstrates that a purely data-driven method can effectively identify the major contributors (e.g., specific SM or IBR plants) to poorly damped low-frequency oscillations. The method requires time-series data from PMUs located only at the point of interconnection (POI) of SM or IBR plants. It is applicable to both natural and forced oscillations and works well for both SM- and IBR-dominated power systems. Using simulated PMU data from a modified IEEE 39-bus system with 100\% IBRs and a 179-bus equivalent of the WECC system, as well as real-world oscillation event data from the ISO New England system, the method is found to be effective across a range of scenarios and test cases. These findings position Extended Dynamic Mode Decomposition (EDMD), grounded in Koopman operator theory, as a promising tool for locating dominant oscillation contributors without requiring extensive PMU coverage

For time-series data with higher nonlinearities (e.g. limit-cycle oscillations), Koopman operator theory suggests that the accuracy of EDMD can be maintained by selecting an appropriate library of observables that captures the relevant dynamics. Developing a systematic procedure for selecting this library of observables will be the focus of future work.

% We will write the conclusion later... 
% \begin{enumerate}
%     \item \textbf{Forced oscillations} – The plant with the highest contribution typically corresponds to the source, which may be a faulty component in a real system or the location where a forcing signal is injected in simulations. Other plants, especially those in close electrical proximity to the source, also tend to exhibit high contributions.
    
%     \item \textbf{Natural oscillations} – If oscillations arise spontaneously due to a prevailing operating condition, the plants with dominant contribution can be pinpointed. Notably, the notion of a source or sink is less well-defined for natural oscillations, as energy is continuously exchanged among contributing plants, with the source/sink role alternating over successive cycles.
% \end{enumerate}

\section*{Acknowledgment}
The authors would thank Prof Mark O'Malley and Prof Janusz Bialek and colleagues in \href{https://www.imperial.ac.uk/uk-electric-power-innovation/}{EPICS global centre}, and \href{https://www.neso.energy/}{NESO} (grid operator in Great Britain) for their feedback.

\ifCLASSOPTIONcaptionsoff
  \newpage
\fi

\bibliographystyle{IEEEtran}
\bibliography{EDMD_paper_refs}

\end{document}